# Charge density wave orders and enhanced superconductivity under pressure in the kagome metal $CsV_3Sb_5$

*Qi Wang[1,2#], Pengfei Kong[1#], Wujun Shi[1,3,4], Cuiying Pei[1], Chenhaoping Wen[1], Lingling Gao[1], Yi Zhao[1], Qiangwei Yin[5], Yueshen Wu[1], Gang Li[1,2], Hechang Lei[5], Jun Li[1,2], Yulin Chen[1,2,6], Shichao Yan[1,2*], Yanpeng Qi[1*]*

[1]*School of Physical Science and Technology, ShanghaiTech University, Shanghai 201210, China*

[2]*ShanghaiTech Laboratory for Topological Physics, ShanghaiTech University, Shanghai 201210, China*

[3] *Center for Transformative Science, ShanghaiTech University, Shanghai 201210, China*

[4] *Shanghai High Repetition Rate XFEL and Extreme Light Facility (SHINE), ShanghaiTech University, Shanghai 201210, China*

[5]*Department of Physics and Beijing Key Laboratory of Opto-electronic Functional Materials & Micro-nano Devices, Renmin University of China, Beijing 100872, China*

[6] *Department of Physics, Clarendon Laboratory, University of Oxford, Parks Road, Oxford OX1 3PU, UK*

*# These authors contributed to this work equally.*

** Correspondence should be addressed to S. C. Y. ( yanshch@shanghaitech.edu.cn); Y. P. Q. (qiyp@shanghaitech.edu.cn);*

## Abstract

**Superconductivity in topological kagome metals has recently received great research interests. Here, charge density wave (CDW) orders and the evolution of superconductivity under various pressures in $CsV_3Sb_5$ single crystal with V kagome lattice are investigated. By using high-resolution scanning tunnelling microscopy /spectroscopy (STM/STS), two CDW orders in $CsV_3Sb_5$ are observed which correspond to 4a×1a and 2a×2a superlattices. By applying pressure, the superconducting transition temperature $T_c$ is significantly enhanced and reaches a maximum value of 8.2 K at**

**around 1 GPa. Accordingly, CDW state is gradually declined as increasing the pressure, which indicates the competing interplay between CDW and superconducting state in this material. The broad superconducting transitions around 0.4 - 0.8 GPa can be related to the strong competition relation among two CDW states and superconductivity. These results demonstrate that $CsV_3Sb_5$ is a new platform for exploring the interplay between superconductivity and CDW in topological kagome metals.**

2D kagome lattice consisting of corner-sharing triangles has been studied deeply for a long time due to frustration-driven exotic quantum phases, such as quantum spin liquid state.[1-3] Moreover, owing to the unique structure, kagome lattice can host dispersionless flat bands due to destructive interference and linearly dispersive Dirac energy bands resemble those in honeycomb lattice.[4,5] When taking spin-orbit coupling and magnetization into consideration, many peculiar phenomena can be observed. For instance, large intrinsic anomalous Hall effect associated with massive Dirac/Weyl points in magnetic kagome metal $Fe_3Sn_2$ and $Co_3Sn_2S_2$,[6-9] negative magnetoresistance related to chiral anomaly and the observations of characteristic surface Fermi-arcs in $Co_3Sn_2S_2$.[9-11] Furthermore, the superconductivity in flat bands system and doped spin-1/2 kagome lattice with breaking time-reversal symmetry has been predicted.[12,13] However, due to the limitation of novel kagome materials, the studies of superconductivity in kagome systems are scarce.

Recently, topological kagome metal family $AV_3Sb_5$ (A = K, Rb, Cs) have attracted significant attention.[14] They have a layered hexagonal centrosymmetric structure with space group *P6/mmm* (No. 191). The crystal structure of $AV_3Sb_5$ is composed of $V_3Sb_5$ slabs and A layers stacking along the *c*-axis (Figure 1a). In the V-Sb slab, the V atoms form 2D kagome layer with one kind of Sb1 atoms occupy the centers of hexagons, and another kind of Sb2 atoms form 2D honeycomb lattice below and above the kagome layer (Figure 1b) . Different from other kagome materials, these compounds exhibit superconductivity, and the superconducting transition temperature $T_c$ is 0.93 K for $KV_3Sb_5$, 0.92 K for $RbV_3Sb_5$, and 2.5 K for $CsV_3Sb_5$, respectively.[15-18] There is also a proximity-induced spin-triplet superconductivity in Josephson Junctions of $K_{1-x}V_3Sb_5$.[19] They undergo anomaly in both magnetization and transport measurements at ≈ 80 K - 110 K.[14-17] Magnetic susceptibility, neutron scattering and muon spin spectroscopy measurements all found no evidence for the existence of long/short range magnetic order and local moments.[14,15,17,20] On the other hand, high-resolution STM on $KV_3Sb_5$ and single crystal x-ray diffraction (XRD) on $CsV_3Sb_5$ strongly suggest the existence of CDW orders.[15,21] Moreover, STM study also found that the topological charge order in $KV_3Sb_5$ shows chiral anisotropy.[21] Furthermore, the results of angle-resolved photoemission spectroscopy (ARPES) and density-functional theoretical calculations indicate that $CsV_3Sb_5$ hosts $Z_2$ topological band structure.[15] Therefore, based on these exotic, rich physical properties, V-based kagome system $AV_3Sb_5$ would open up a new platform to investigate the interplay between superconductivity, CDW and nontrivial topological electronic structure. Generally, applying pressure on materials is a powerful, clear tool to study the competing relationship between the superconductivity and CDW.[22-24]

In this work, we perform STM/STS measurements and demonstrate that the coexistence of $4\mathbf{a}\times1\mathbf{a}$ and $2\mathbf{a}\times2\mathbf{a}$ CDW orders at low temperature in $CsV_3Sb_5$ single crystal with superconducting transition temperature ≈ 2.5 K. Upon application of pressure, the superconducting transition temperature is enhanced significantly to a maximum value ≈ 8.2 K at around 1 GPa, a dome-shaped phase diagram was observed accompanying the suppression of CDW state. Our experiments reveal a complex landscape of electronic states that can co-exist on a kagome lattice and compete with superconductivity, and offer more insights into the rich physics related to the electronic correlations in this novel family of V-based topological kagome metals.

The longitudinal resistivity $\rho_{xx}$ ($T$) as a function of temperature at zero field is shown in Figure 1e. It shows metallic behavior at normal state and an anomaly at $T_{CDW}$ ≈ 94 K, indicating the appearance of CDW state. There is a remarkably drop in the $\rho_{xx}$ ($T$) curve at low temperature (inset of Figure 1e), corresponding to the superconductivity transition in $CsV_3Sb_5$ single crystal. The onset superconductivity transition temperature $T_{c,onset}$ and zero resistivity temperature $T_{c,zero}$ is about 3.46 and 2.5 K, respectively, in good agreement with previous results.[15] As shown in Figure 1f, the magnetic susceptibility $\chi$ ($T$) curves with zero-field-cooling (ZFC) and field-cooling (FC) modes at $\mu_0$H = 1T for $H$ || $c$ do not show any signature of magnetic order in the whole temperature range. In addition, we observe a sharp decline around 94 K in $\chi$ ($T$) curve, further indicating the existence of CDW state. The temperature dependence of magnetic susceptibility $4\pi\chi$ ($T$) with ZFC and FC modes at 1 mT for $H$ || $ab$ is shown in the inset of Figure 1f. The determined $T_c$ for $CsV_3Sb_5$ single crystal is about 2.49 K, very close to the value derived from $\rho_{xx}$ ($T$) curve.

To gain further insight into the CDW orders and superconductivity in $CsV_3Sb_5$, we performed low-temperature STM measurements on the $CsV_3Sb_5$ single crystal. As shown in Figure 1c,d, there are two typical cleaving surfaces with the Cs layer and the Sb layer, respectively. The superconducting gap can be measured on both these surfaces and can be suppressed by external magnetic field (Figure 1g,h). In the Sb-terminated surface, there are the randomly distributed Cs atoms, and we can also find large Cs-atom-free regions for imaging. Figure 2a shows the high-resolution STM topography taken on the Cs-atom-free region in the Sb-terminated surface. Around the intrinsic defects in the Sb surface, a series of concentric circles can be observed, they are the quasiparticle-interference (QPI) patterns of the electronic states around Γ point in $CsV_3Sb_5$ (Figure S2, Supporting Information). The superstrctures and stripe-like superlattices can also been seen in Figure 2a. Their periodicities can be indentified in the Fourier-transform (FT) image of the STM topography (Figure 2b, c) The 2**a**×2**a** superstructure is a CDW order and it is related to the the magnetization and transport anomaly at $T_{CDW}$ ≈ 94 K.[15,25-27] This 2**a**×2**a** CDW order is driven by the Peierls instability characterized by the softening of a breathing phonon mode in the kagome lattice (Figure S3, Supporting Information).[28] This calculated in-plane component of CDW order (2**a**×2**a** superstructure) is consistent with the STM results. It is noted that recent experimental studies indicate there is also a modulation along the *c* axis, that is, 3D CDW order with 2**a**×2**a**×2**c** superlattice.[27,29,30] In addtion to 2**a**×2**a** pattern, the STM topography also shows the stripe-like pattern which has 4**a**×1**a** periodicity (Figure 2b,c). This stripe-like superstructure has not been experimentally observed in the cousin compound $KV_3Sb_5$ and other kagome systems. Its origin still needs further investigations. The d$I$ / d$V$ spectra taken on the defect-free region (Figure 2d) are spatially uniform. According to the band calculations (Figure 2e), the −220 meV peak in the d$I$ / d$V$ spectra corresponds to the van Hove singularity (VHS) of the electronic states at **M** point, and the dip at −300 meV is related to the Dirac point (DP) in $CsV_3Sb_5$.[26,31] We note that there is an energy dip located near the Fermi level which may be the CDW gap in this material.

In order to understand the relationship between superconductivity and CDW, we study the temperature dependence of electrical resistivity $\rho(T)$ at various pressures using 7373 as pressure transmitting medium (PTM; Figure 3a). They all show metallic behavior in the whole pressure range. With applying pressure, we found that CDW temperature $T_{\mathrm{CDW}}$ gradually shifts to lower temperature (Figure S4, Supporting Information). Here, we only observe the change of CDW transition with 2**a**×2**a** superlattice under pressure in transport measurement. The absence of 4**a**×1**a** CDW transition in $\rho(T)$ curves possibly be related to the small gaped Fermi surfaces (FSs) along $\boldsymbol{Q}_{4a}$ vector. In contrast, the $T_{\mathrm{c,onset}}$ increases continuously with increasing pressure and reaches a maximum value ≈ 8.0 K at 0.82 GPa (Figure 3b). Interestingly, the superconducting transition width $\Delta T_{\mathrm{c}}$ at 0.41 and 0.82 GPa becomes large and reaches 2.06 K and 2.43 K, respectively. In addition, we have also performed the high-pressure measurements on $CsV_3Sb_5$ single crystal in three independent runs using NaCl as PTM (Figure 3c; Figure S5, Supporting Information). These results show that our experiment is reproducible and reliable. Interestingly, there is a reentering superconductivity when the pressure above 15.32 GPa (Figure 3c; Figures S5 and S6, Supporting Information), the two superconducting domes are also observed in other works, such as $LaFeAsO_{1-x}H_x$ (x<0.53) and $CeCu_2Si_2$ etc.[32-36] $\rho(T)$ as a function of temperature at various fields in different runs is presented in Figure S5, Supporting Information. It can be clearly seen that $T_c$ is gradually suppressed by magnetic field. Deviating from the Werthamer–Helfand–Hohenberg theory based on the single-band model, the upper critical field, $\mu_0H_{c2}(T)$, has a positive curvature close to $T_c$ ($H$ = 0 T), as shown in Figure 3d. This is similar to the behaviors of both $MoTe_2$[37] and 2H-$NbSe_2$.[38] The upper critical field $\mu_0H_{c2}(T)$ curves in different runs can be well fitted by the formula $\mu_0H_{c2}(T) = \mu_0H_{c2}(0)[1-(T/T_c)]^{1+\alpha}$, where $\mu_0H_{c2}(0)$ is the upper critical field at $T$ = 0 K, $T_c$ is determined at the 50% drop of normal state resistivity. The fitted $\mu_0H_{c2}(0)$ is 6.4 T at 1.07 GPa, which yields a Ginzburg–Landau coherence length $\xi_{GL}(0)$ [ $=\sqrt{\Phi_0/2\pi\mu_0H_{c2}(0)}$ ] of ≈ 7 nm. The somewhat higher $\mu_0H_{c2}(0)$ indicates the high sample quality of the as-grown $CsV_3Sb_5$. The corresponding data obtained at 1.18, 4.34, and 50.71 GPa is also shown in Figure 3d. It should be noted that the $\mu_0H_{c2}(0)$ in different runs is smaller than Pauli limiting field $H_P(0)$ ($1.84T_c$ = 8.1, 14.7, 13.87, and 7.6 T), suggesting that the Pauli paramagnetic effect could be negligible.

As shown in Figure 3f, the combined phase diagram of $T_{CDW}$ and $T_c$ provides a clear relation between CDW and superconductivity. The superconductivity coexists with CDW in $CsV_3Sb_5$ under ambient pressure. Below 1 GPa, $T_c$ is very sensitive to pressure and increases remarkably with applying pressure, accompanied by a decline of $T_{CDW}$. This observation suggests that there is a mutually competing interaction between CDW and superconductivity, that is, enhanced $T_c$ is accompanied by strong suppression of CDW order under pressure. Raman spectroscopy measurements found no evidence for the structural phase transition below 10.84 GPa (Figure 3e; Figure S7, Supporting Information), demonstrating the increased $T_c$ at low pressure range cannot arise from structural phase transition. It might be a consequence of the variation in the density of states at the Fermi level under pressure. At ambient pressure, the existence of two CDW states in $CsV_3Sb_5$ results in the gap opening over part of the FSs in the direction of the $\boldsymbol{Q_{2a}}$ and $\boldsymbol{Q_{4a}}$ vectors. As pressure increases, both of two CDW states are gradually suppressed and the FSs could be restored, accordingly $T_c$ increases. In addition, it should be remarked that $T_c$ continues to increase after the absence of $T_{CDW}$ and its maximum value $T_{c,onset}$ reaches ≈ 8.2 K at around 1 GPa. It could be due to the CDW state along $\boldsymbol{Q_{4a}}$ vector direction is not destructed completely by pressure when the 2**a**×2**a** CDW state is fully suppressed, leading to the further increase of $T_c$ until the 4**a**×1**a** CDW state disappeared above 1 GPa. In recent studies,[39] the disappearance of the CDW states could be related to the suppression of $Sb_2$ atoms displacements which is crucial for the stability of CDW states, according to the results of anisotropic compression in high-pressure XRD and the unperturbed electronic structure under this pressure range. On the other hand, the values of $\Delta T_c$ at 0.4 - 0.8 GPa become larger than those in other pressure ranges, which is also observed in recent study.[40-41] It might be attributed to the extremely strong competitive interplay among CDW states with 4**a**×1**a** and 2**a**×2**a** superlattices and superconductivity in $CsV_3Sb_5$. When the pressure is above 1 GPa, $T_{c,onset}$ starts to decrease consecutively and reaches 2.07 K at 10.3 GPa. Finally, it disappears completely with pressure up to 13.13 GPa (Figure 3f).

In summary, we successfully grow the $CsV_3Sb_5$ single crystal using the self-flux method. With applying pressure, the CDW transition at ≈ 94 K under ambient pressure is suppressed gradually and the $T_c$ enhances significantly up to 8.2 K. The observation of two CDW states with 4**a**×1**a** and 2**a**×2**a** superlattices by STM experiment indicates that the broad superconducting transition could be related to the strong competitive relationship between CDW states and superconductivity. Recently theoretical calculation indicates that the nontrivial $Z_2$ -type topological band structure remains the same after the CDW transition. The electron-phonon coupling in the CDW phase is too weak to rationalize the superconductivity in $CsV_3Sb_5$.[28] Thus, the resistivity anomaly/CDW, a nontrivial topological state, and possible unconventional superconductivity are all observed in $CsV_3Sb_5$, all contributing to the highly interesting physics seen in this in V-based kagome metals. Future experiments should address the relation of different phases by more detailed measurements, while also searching for evidence of intrinsic topological superconductivity and Majorana modes.

***Note added***. When we prepared the manuscript, we learned that similar work was carried out independently by another group and published as an e-print.[26] The observation of two CDW states with 4**a**×1**a** and 2**a**×2**a** superlattices in that paper is consistent with our results.

**Experimental Section**

*Sample synthesis and characterization* : $CsV_3Sb_5$ single crystals were grown by Sb flux.[14] The XRD pattern of a single crystal was performed using a Bruker D8 X-ray diffractometer with Cu $K_\alpha$ radiation ($\lambda$ = 0.15418 nm) at room temperature.

*Transport, magnetization and Raman spectroscopy*: Magnetization and electrical transport measurements were carried out by using quantum design magnetic property measurement system (MPMS3) and physical property measurement system (PPMS-9), respectively. The longitudinal resistivity at ambient pressure was measured by using a standard four-probe method. *In situ* high-pressure Raman spectroscopy of single crystal $CsV_3Sb_5$ was measured in Raman spectrometer (Renishaw inVia, U.K.) with laser excitation wavelength $\lambda$ (532 nm) and low-wavenumber filter at room temperature. *In situ* high-pressure resistivity of single crystal $CsV_3Sb_5$ was performed using van der Pauw four-probe method in a nonmagnetic diamond anvil cell (DAC) with anvil culet sizes of 400 μm. The Cubic BN/epoxy mixture as the insulating material was placed between BeCu gasket and Pt electrical leads. Daphne 7373 and NaCl was used as PTM in different runs, respectively. The values of pressure in DAC were determined by measuring the luminescence for small ruby chips on the surface of sample at room temperature.[42]

*High-resolution scanning tunnelling microscopy /spectroscopy (STM/STS):* STM experiments were carried out with a Unisoku low-temperature STM at a base temperature of 4.3 K (STS shown in Figure 1g,h were conducted with a home-built He-3 STM working at 0.6 K). The $CsV_3Sb_5$ single crystals were cleaved at 77 K under ultrahigh vacuum, and after that the samples were transferred into the STM for measurements. Unless otherwise specified, STS were done by using standard lock-in technique with 3 mV modulation at the frequency of 914 Hz.

*Theoretical calculations*: The first principle calculations were performed based on density functional theory as implemented in Vienna *Ab initio* Simulation Package (VASP) [43] with the projector augmented wave potential.[44,45] The exchange-correlation potential was formulated by generalized gradient approximation with Perdew-Burke-Ernzerhof functional.[46] A Γ-center 10 × 10 × 6 *k* points grid was used for the first Brillouin zone sampling. In order to determine the CDW order, the phonon dispersion was carried out using the finite displacement method with VASP and the PHONOPY package.[47] A 2 × 2 × 1 inverse Star of David structure was employed.[28]

**Acknowledgements**

This work was supported by the National Key R&D Program of China (Grant No. 2018YFA0704300, 2018YFE0202600 and 2016YFA0300504), the National Natural Science Foundation of China (Grant No. U1932217, 11974246, 11822412, 11774423, 11874042, 12004250 and 61771234), the Natural Science Foundation of Shanghai (Grant No. 19ZR1477300), the Science and Technology Commission of Shanghai Municipality (19JC1413900 and 20YF1430700), the Beijing Natural Science Foundation (Grant No. Z200005) and the China Postdoctoral Science Foundation (Grant No. 2021M692132). The authors thank the support from C*ħ*EM (02161943) and Analytical Instrumentation Center (SPST-AIC10112914), SPST, ShanghaiTech University.

## References

[1] L. Balents, *Nature* **2010**, *464*, 199.

[2] T. H. Han, J. S. Helton, S. Chu, D. G. Nocera, J. A. Rodriguez-Rivera, C. Broholm, Y. S. Lee, *Nature* **2012**, *492*, 406.

[3] Y. Zhou, K. Kanoda, T.-K. Ng, *Rev. Mod. Phys.* **2017**, *89*, 025003.

[4] Z. H. Liu, M. Li, Q. Wang, G. Wang, C. Wen, K. Jiang, X. Lu, S. Yan, Y. Huang, D. Shen, J.-X. Yin, Z. Wang, Z. P. Yin, H. C. Lei, S. C. Wang, *Nat. Commun.* **2020**, *11*, 4002.

[5] M. Kang, L. Ye, S. Fang, J.-S. You, A. Levitan, M. Han, J. I. Facio, C. Jozwiak, A. Bostwick, E. Rotenberg, M. K. Chan, R. D. McDonald, D. Graf, K. Kaznatcheev, E. Vescovo, D. C. Bell, E. Kaxiras, J. van den Brink, M. Richter, M. P. Ghimire, J. G. Checkelsky, R. Comin, *Nat. Mater.* **2020**, *19*, 163.

[6] Q. Wang, S. S. Sun, X. Zhang, F. Pang, H. C. Lei, *Phys. Rev. B* **2016**, *94*, 075135.

[7] L. Ye, M. Kang, J. Liu, F. von Cube, C. R. Wicker, T. Suzuki, C. Jozwiak, A. Bostwick, E. Rotenberg, D. C. Bell, L. Fu, R. Comin, J. G. Checkelsky, *Nature* **2018**, *555*, 638.

[8] Q. Wang, Y. Xu, R. Lou, Z. Liu, M. Li, Y. Huang, D. Shen, H. Weng, S. C. Wang, H. C. Lei, *Nat. Commun.* **2018**, *9*, 3681.

[9] E. Liu, Y. Sun, N. Kumar, L. Muechler, A. Sun, L. Jiao, S.-Y. Yang, D. Liu, A. Liang, Q. Xu, J. Kroder, V. Süβ, H. Borrmann, C. Shekhar, Z. Wang, C. Xi, W. Wang, W. Schnelle, S. Wirth, Y. Chen, S. T. B. Goennenwein, C. Felser, *Nat. Phys.* **2018**, *14*, 1125.

[10] D. F. Liu, A. J. Liang, E. K. Liu, Q. N. Xu, Y. W. Li, C. Chen, D. Pei, W. J. Shi, S. K. Mo, P. Dudin, T. Kim, C. Cacho, G. Li, Y. Sun, L. X. Yang, Z. K. Liu, S. S. P. Parkin, C. Felser, Y. L. Chen, *Science* **2019**, *365*, 1282.

[11] N. Morali, R. Batabyal, P. K. Nag, E. Liu, Q. Xu, Y. Sun, B. Yan, C. Felser, N. Avraham, H. Beidenkopf, *Science* **2019**, *365*, 1286.

[12] S. Miyahara, S. Kusuta, N. Furukawa, *Physica C* **2007**, *460*, 1145.

[13] W.-H. Ko, P. A. Lee, X.-G. Wen, *Phys. Rev. B* **2009**, *79*, 214502.

[14] B. R. Ortiz, L. C. Gomes, J. R. Morey, M. Winiarski, M. Bordelon, J. S. Mangum, I. W. H. Oswald, J. A. Rodriguez-Rivera, J. R. Neilson, S. D. Wilson, E. Ertekin, T. M. McQueen, E. S. Toberer, *Phys. Rev. Mater.* **2019**, *3*, 094407.

[15] B. R. Ortiz, S. M. L. Teicher, Y. Hu, J. L. Zuo, P. M. Sarte, E. C. Schueller, A. M. M. Abeykoon, M. J. Krogstad, S. Rosenkranz, R. Osborn, R. Seshadri, L. Balents, J. He, S. D. Wilson, *Phys. Rev. Lett.* **2020**, *125*, 247002.

[16] B. R. Ortiz, P. M. Sarte, E. M. Kenney, M. J. Graf, S. M. L. Teicher, R. Seshadri, S. D. Wilson, *Phys. Rev. Mater.* **2021**, *5*, 034801.

[17] Q. W. Yin, Z. J. Tu, C. S. Gong, Y. Fu, S. H. Yan, H. C. Lei. *Chin. Phys. Lett.* **2021**, *38*, 037403.

[18] S. Ni, S. Ma, Y. Zhang, J. Yuan, H. Yang, Z. Lu, N. Wang, J. Sun, Z. Zhao, D. Li, S. Liu, H. Zhang, H. Chen, K. Jin, J. Cheng, L. Yu, F. Zhou, X. Dong, J. Hu, H.-J. Gao, Z. X Zhao, *Chin. Phys. Lett.* **2021**, *38*, 057403.

[19] Y. Wang, S.-Y. Yang, P. K. Sivakumar, B. R. Ortiz, S. M. L. Teicher, H. Wu, A. K. Srivastava, C. Garg, D. Liu, S. S. P. Parkin, E. S. Toberer, T. McQueen, S. D. Wilson, M. N. Ali, arXiv: 2012.05898 **2020**.

[20] E. M. Kenney, B. R. Ortiz, C. Wang, S. D. Wilson, M. J. Graf, *J. Phys.: Condens. Matter* **2021**, *33*, 235801.

[21] Y.-X. Jiang, J.-X. Yin, M. M. Denner, N. Shumiya, B. R. Ortiz, G. Xu, Z. Guguchia, J. He, M. S. Hossain, X. Liu, J. Ruff, L. Kautzsch, S. S. Zhang, G. Chang, I. Belopolski, Q. Zhang, T. A. Cochran, D. Multer, M. Litskevich, Z.-J. Cheng, X. P. Yang, Z. Q. Wang, R. Thomale, T. Neupert, S. D. Wilson, M. Z. Hasan,, *Nat. Mater.* **2021**, *10*, 1038.

[22] A. F. Kusmartseva, B. Sipos, H. Berger, L. Forró, E. Tutiš, *Phys. Rev. Lett.* **2009**, *103*, 236401.

[23] Q.-G. Mu, D. Nenno, Y.-P. Qi, F.-R. Fan, C. Pei, M. ElGhazali, J. Gooth, C. Felser, P. Narang, S. Medvedev, arXiv:2010.07345 **2020**.

[24] Y. P. Qi, W. J. Shi, P. G. Naumov, N. Kumar, W. Schnelle, O. Barkalov, C. Shekhar, H. Borrmann, C. Felser, B. Yan, S. A. Medvedev, *Phys. Rev. B* **2016**, *94*, 054517.

[25] C. C. Zhao, L. S. Wang, W. Xia, Q. W. Yin, J. M. Ni, Y. Y. Huang, C. P. Tu, Z. C. Tao, Z. J. Tu, C. S. Gong, H. C. Lei, Y. F. Guo, X. F. Yang, S. Y. Li, arXiv:2102.08356 **2021**.

[26] H. Zhao, H. Li, B. R. Ortiz, S. M. L. Teicher, T. Park, M. Ye, Z. Wang, L. Balents, S. D. Wilson, I. Zeljkovic, arXiv:2103.03118 **2021**.

[27] Z. Liang, X. Hou, W. Ma, F. Zhang, P. Wu, Z. Zhang, F. Yu, J. J. Ying, K. Jiang, L. Shan, Z. Y. Wang, X. H. Chen, arXiv:2103.04760 **2021**.

[28] H. X. Tan, Y. Z. Liu, Z. Q. Wang, B. H. Yan, *Phys. Rev. Lett.* **2021**, *127*, 046401.

[29] H. X. Li, T. T. Zhang, Y. Y. Pai, C. Marvinney, A. Said, T. Yilmaz, Q. Yin, C. Gong, Z. Tu, E. Vescovo, R. G. Moore, S. Murakami, H. C. Lei, H. N. Lee, B. Lawrie, H. Miao, arXiv:2103.09769 **2021**.

[30] D. W. Song, L. X. Zheng, F. H. Yu, J. Li, L. P. Nie, M. Shan, D. Zhao, S. J. Li, B. L. Kang, Z. M. Wu, Y. B. Zhou, K. L. Sun, K. Liu, X. G. Luo, Z. Y. Wang, J. J. Ying, X. G. Wan, T. Wu, X. H. Chen, arXiv:2104.09173 **2021.**

[31] H. Chen, H. Yang, B. Hu, Z. Zhao, J. Yuan, Y. Xing, G. Qian, Z. Huang, G. Li, Y. Ye, Q. W. Yin, C. S. Gong, Z. J. Tu, H. C. Lei, S. Ma, H. Zhang, S. Ni, H. Tan, C. Shen, X. Dong, B. H. Yan, Z. Q. Wang, H.-J. Gao, arXiv:2103.09188 **2021**.

[32] S. Iimura, S. Matsuishi, H. Sato, T. Hanna, Y. Muraba, S. W. Kim, J. E. Kim, M. Takata, H. Hosono, *Nat. Commun.* **2012**, *3*, 943.

[33] H. Q. Yuan, F. M. Grosche, M. Deppe, C. Geibel, G. Sparn, F. Steglich, *Science* **2003**, *302*, 2104.

[34] Z. Zhang, Z. Chen, Y. Zhou, Y. Yuan, S. Wang, J. Wang, H. Yang, C. An, L. Zhang, X. Zhu, Y. Zhou, X. Chen, J. Zhou, Z. R. Yang, *Phys. Rev. B*, **2021**, *103*, 224513.

[35] X. Chen, X. Zhan, X. Wang, J. Deng, X.-B. Liu, X. Chen, J.-G. Guo, X. L. Chen, *Chin. Phys. Lett.* **2021**, *38*, 057402.

[36] C. C. Zhu, X. Yang, W. Xia, Q. Yin, L. S. Wang, C. C. Zhao, D. Z. Dai, C. Tu, B. Q. Song, Z. Tao, Z. Tu, C. Gong, H. Lei, Y. F. Guo, S. Y. Li, arXiv:2104.14487 **2021**.

[37] Y. P. Qi, P. G. Naumov, M. N. Ali, C. R. Rajamathi, W. Schnelle, O. Barkalov, M. Hanfland, S.-C. Wu, C. Shekhar, Y. Sun, V. Süß, M. Schmidt, U. Schwarz, E. Pippel, P. Werner, R. Hillebrand, T. Föster, E. Kampert, S. Parkin, R. J. Cava, C. Felser, B. H. Yan, S. A. Medvedev. *Nat. Commun.* **2016**, *7*, 11038.

[38] H. Suderow, V. G. Tissen, J. P. Brison, J. L. Martínez, S. Vieira. *Phys. Rev. Lett.* **2005**, *95*, 117006.

[39] A. A. Tsirlin, P. Fertey, B. R. Ortiz, B. Klis, V. Merkl, M. Dressel, S. D. Wilson, E. Uykur, arXiv:2105.01397 **2021**.

[40] K. Y. Chen, N. N. Wang, Q. W. Yin, Y. H. Gu, K. Jiang, Z. J. Tu, C. S. Gong, Y. Uwatoko, J. P. Sun, H. C. Lei, J. P. Hu, J.-G. Cheng. *Phys. Rev. Lett.* **2021**, *126*, 247001.

[41] F. H. Yu, D. H. Ma, W. Z. Zhuo, S. Q. Liu, X. K. Wen, B. Lei, J. J. Ying, X. H. Chen, *Nat. Commun.* **2021**, *12*, 3645.

[42] H. K. Mao, J. Xu, P. M. Bell, *J. Geophys. Res.* **1986,** *91*, 4673.

[43] G. Kresse, J. Furthmüller, *Phys. Rev. B* **1996**, *54*, 11169.

[44] P. E. Blöchl, *Phys. Rev. B*, **1994**, *50*, 17953.

[45] G. Kresse and D. Joubert, *Phys. Rev. B*, **1999**, *59*, 1758.

[46] J. P. Perdew, K. Burke, M. Ernzerhof, *Phys. Rev. Lett.* **1996**, **77**, 3865.

[47] A. Togo and I. Tanaka, *Scr. Mater.* **2015**, *108*, 1.

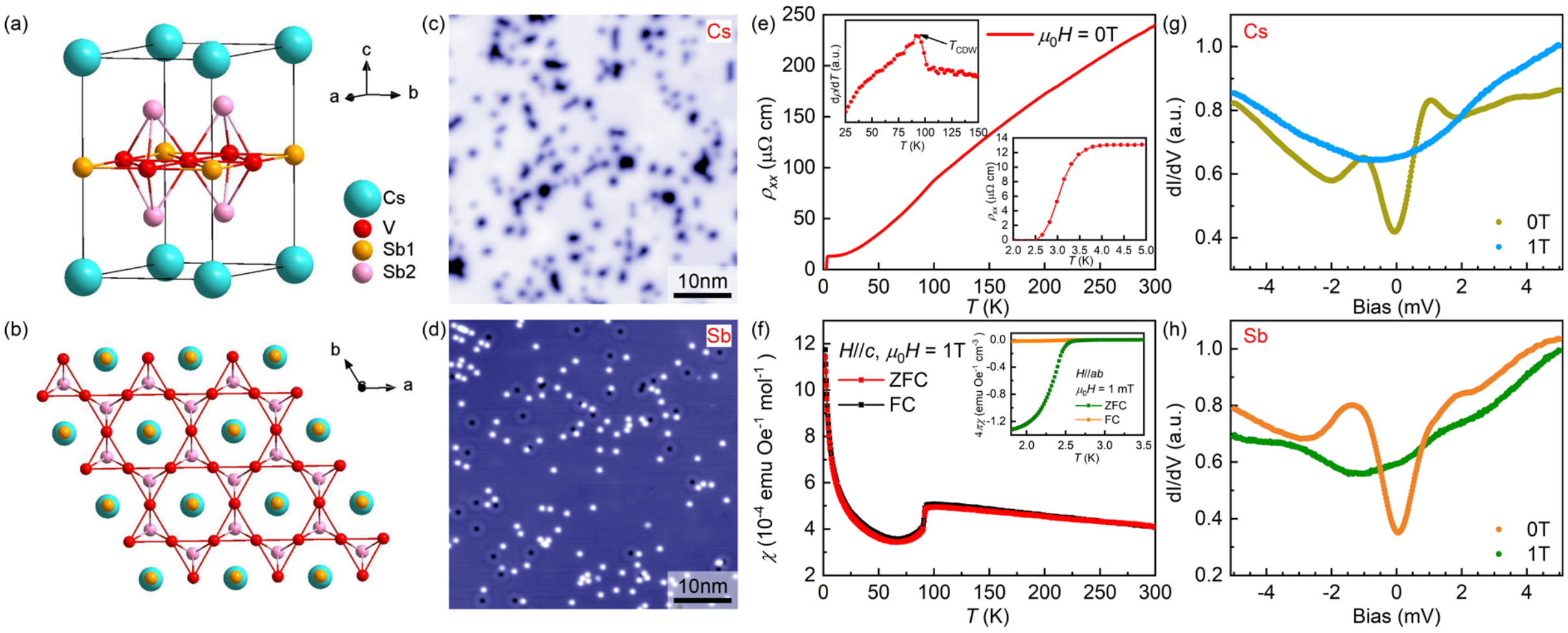

**Figure 1.** Structure, STM topographies, longitudinal resistivity, magnetic susceptibility, and d$I$/d$V$ spectra of $CsV_3Sb_5$ single crystal. a) Crystal structure of $CsV_3Sb_5$ and b) the top view of Cs layer and V-Sb slabs. c, d) Constant-current STM topographies taken on the Cs- and Sb-terminated surfaces, respectively ($V$s = 500 mV, $I$ = 20 pA; $V$s = −500 mV, $I$ = 20 pA). e) Temperature dependence of longitudinal resistivity $\rho_{xx}$ ($T$) at zero field. The insets are enlarged view of $\rho_{xx}$ ($T$) below 5 K and d$\rho$/d$T$ versus $T$ curve. The position of $T_{CDW}$ is marked by arrow. f) Magnetic susceptibility $\chi$ ($T$) as a function of temperature with ZFC and FC modes at $\mu_0H$ = 1T for $H$ // $c$. Inset: temperature dependence of magnetic susceptibility 4π$\chi$ ($T$) with ZFC and FC modes at $\mu_0H$ = 1 mT for $H$ // $ab$. g, h) d$I$ /d $V$ spectra measured without and with 1 T magnetic field on the Cs- and Sb-terminated surfaces, respectively. The d$I$ /d $V$ spectra in (g, h) are measured at 0.6 K with 50 μV modulation.

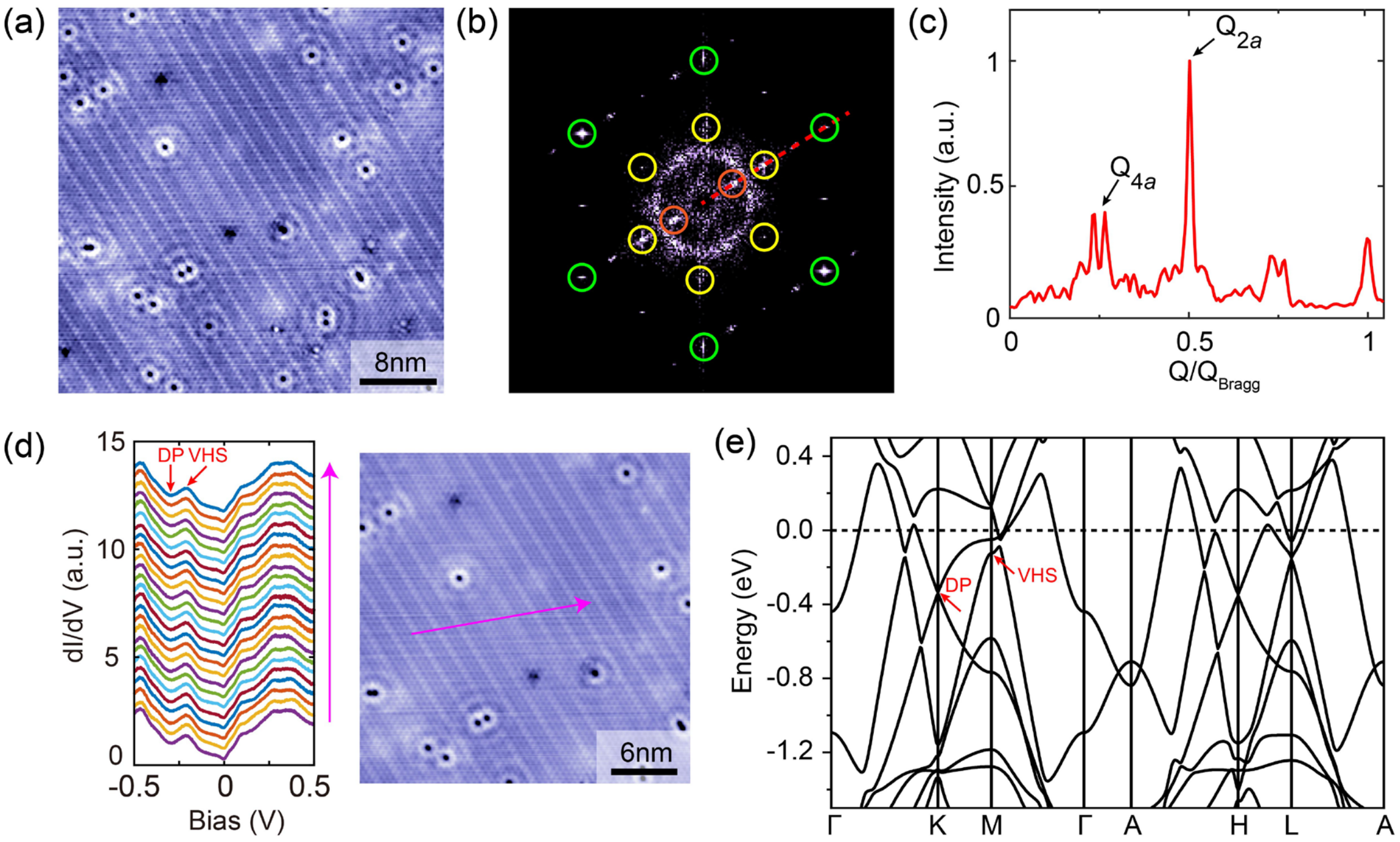

**Figure 2.** STM and first principle calculations of $CsV_3Sb_5$. a) High-resolution STM topography taken on the Sb-terminated surface with $V$s = 100 mV, $I$ = 200 pA. b) Fourier transform (FT) of the topography shown in panel (a). The Qbragg peaks, 2**a**×2**a** and 4**a**×1**a** CDW vectors are marked by the green, yellow and orange circles, respectively. c) Line profile along the red dashed line shown in (b). d) d$I$/d$V$ spectra (left) taken along the purple arrow shown in the STM topography (right). The spectra are vertically offset for clarity. e) The band structure of $CsV_3Sb_5$, van Hove singularity (VHS) at **M** point and Dirac point (DP) at $k$ point are indicated by the red arrows.

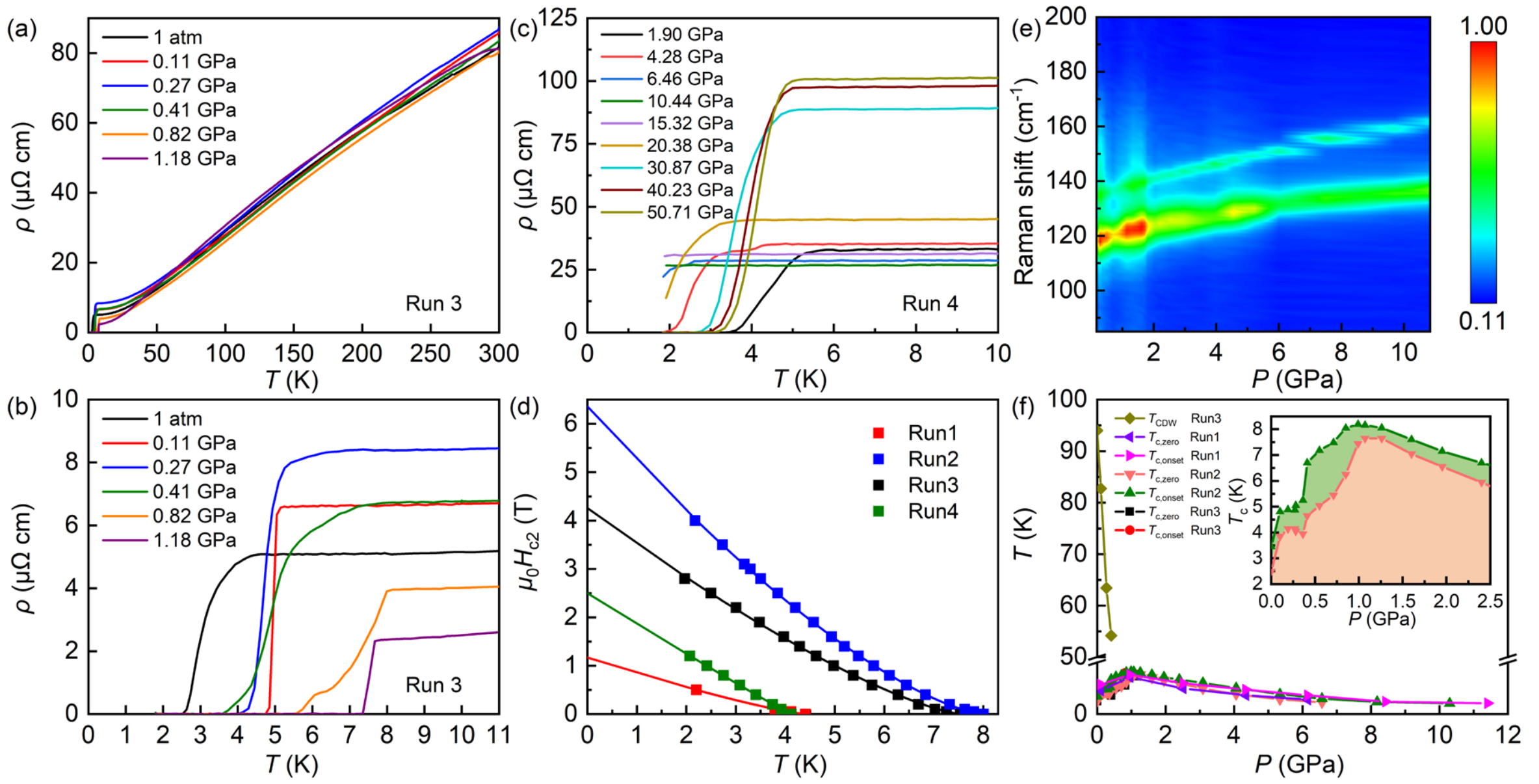

**Figure 3.** CDW orders and the evolution of superconductivity under various pressures in $CsV_3Sb_5$ single crystal. a) Temperature dependence of resistivity $\rho(T)$ at different pressures in run 3 using 7373 as PTM. b) Enlarged view of the superconducting transition at low temperatures in run 3. c) Enlarged part of $\rho(T)$ at low temperature region in run 4 using NaCl as PTM. d) Upper critical field $\mu_0H_{c2}(T)$ as a function of temperature in different runs, corresponding to the 50% drop of the normal state resistivity. The $\mu_0H_{c2}(T)$ curves for run1, run2, run 3 and run 4 were measured at 4.34, 1.07, 1.18. and 50.71 GPa, respectively. The solid lines represent the fitting curves using the formula $\mu_0H_{c2}(T) = \mu_0H_{c2}(0)[1-(T/T_c)]^{1+\alpha}$, $\alpha$ represents the positive curvature of $\mu_0H_{c2}(T)$. e) Raman shift for $CsV_3Sb_5$ in compression. f) The temperature-pressure phase diagram of $CsV_3Sb_5$ in different runs. $T_{CDW}$ and $T_c$ ($T_{c,zero}$ and $T_{c,onset}$) are determined from $\rho(T)$ curves. Inset: pressure dependence of the $T_c$ ($T_{c,zero}$ and $T_{c,onset}$) below 2.5 GPa in run 2.

Supporting Information

**Charge density wave orders and enhanced superconductivity under pressure in the kagome metal $CsV_3Sb_5$**

*Qi Wang[1,2#], Pengfei Kong[1#], Wujun Shi[1,3,4], Cuiying Pei[1], Chenhaoping Wen[1], Lingling Gao[1], Yi Zhao[1], Qiangwei Yin[5], Yueshen Wu[1], Gang Li[1,2], Hechang Lei[5], Jun Li[1,2], Yulin Chen[1,2,6], Shichao Yan[1,2*], Yanpeng Qi[1*]*

**Experimental Section**

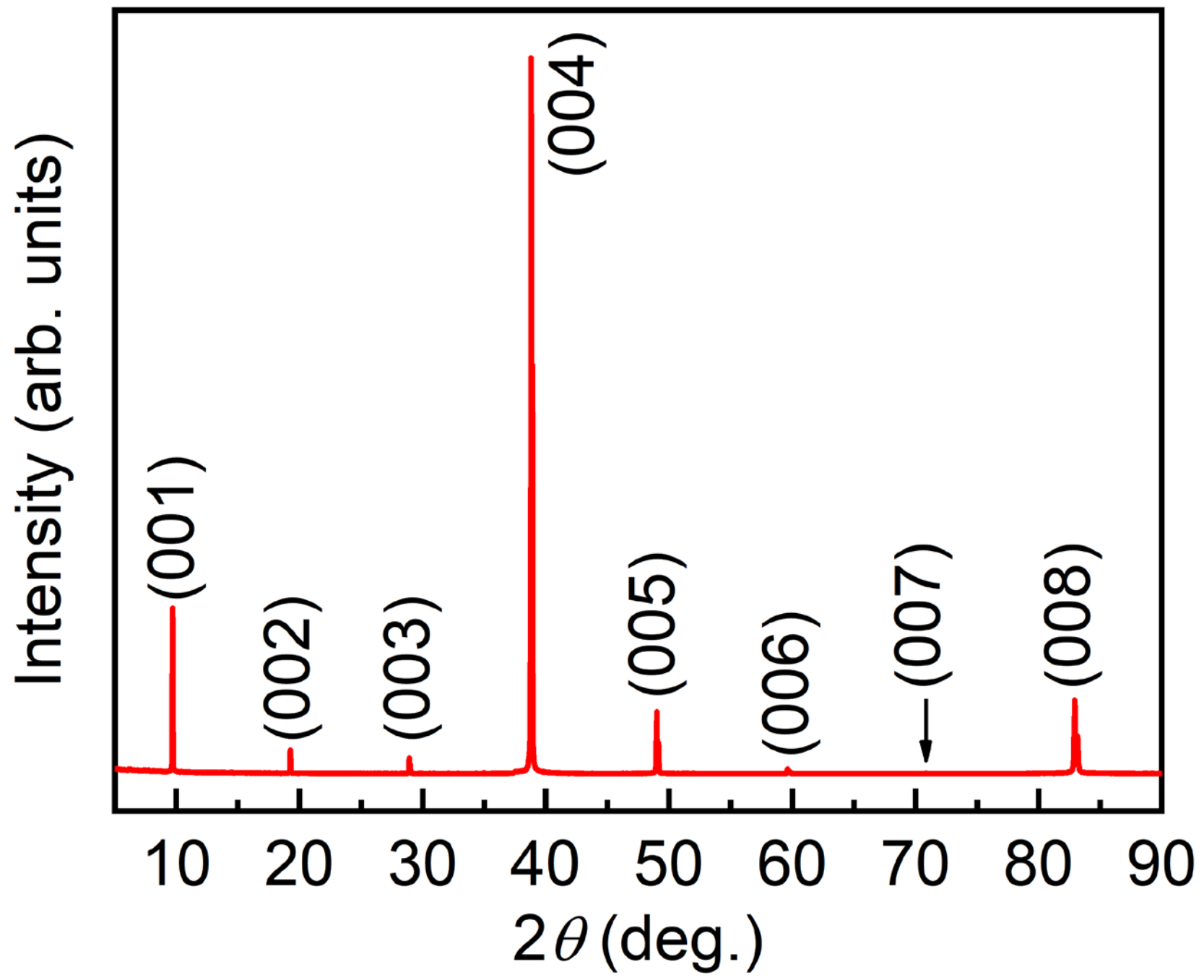

**Figure S1.** XRD pattern of a $CsV_3Sb_5$ single crystal. It indicates that the surface of $CsV_3Sb_5$ single crystal is the (00*l*) plane.

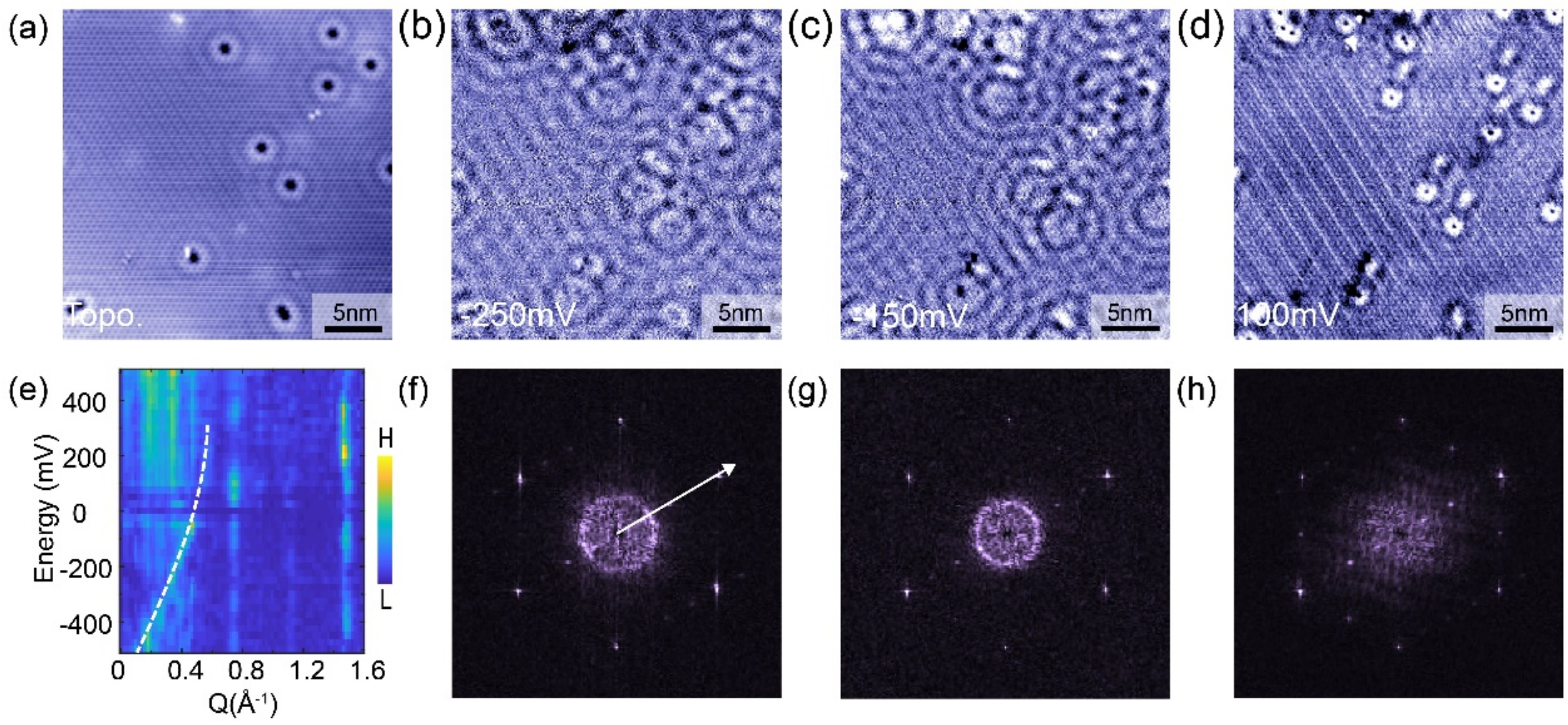

**Figure S2.** a) Constant-current STM topography taken on Sb surface with $V_s = -500$ mV, $I =$ 1 nA. b-d) d$I$ / d$V$ maps taken on the same region shown in (a) with −250 mV (b), −150 mV (c), 100 mV (d) bias voltages. e) Line profile of the FT along the arrow shown in panel (f) as a function of energy, showing the dispersive energy band. f-h) Fourier transform images of the d$I$ / d$V$ maps shown in (b)-(d).

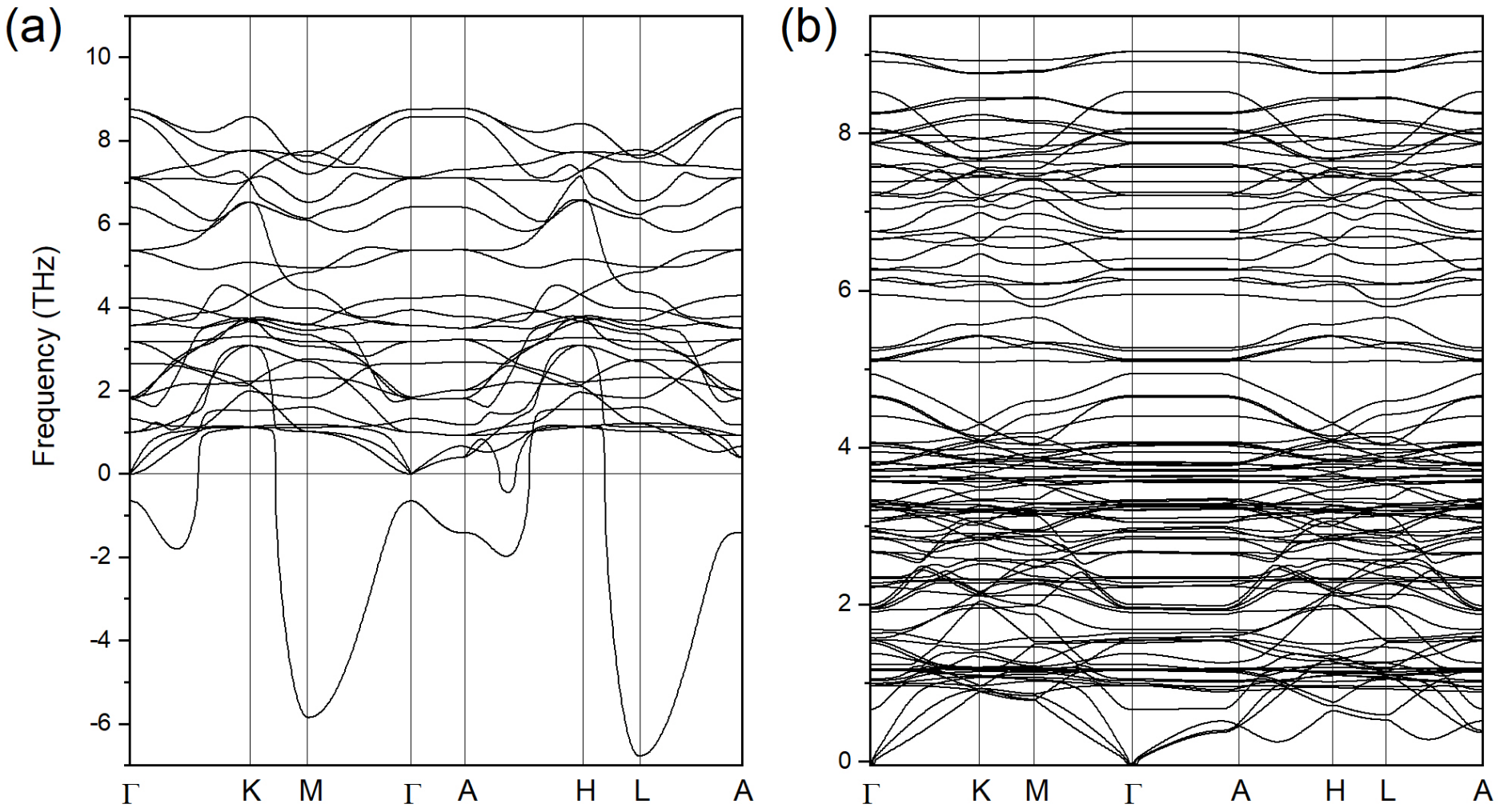

**Figure S3.** The calculated phonon spectra of $CsV_3Sb_5$ without (a) and with (b) the CDW order. A 2 × 2 × 1 Inverse Star of David structure was employed. No imaginary frequency was found for the calculated phonon spectra with the CDW, indicating dynamical stability.

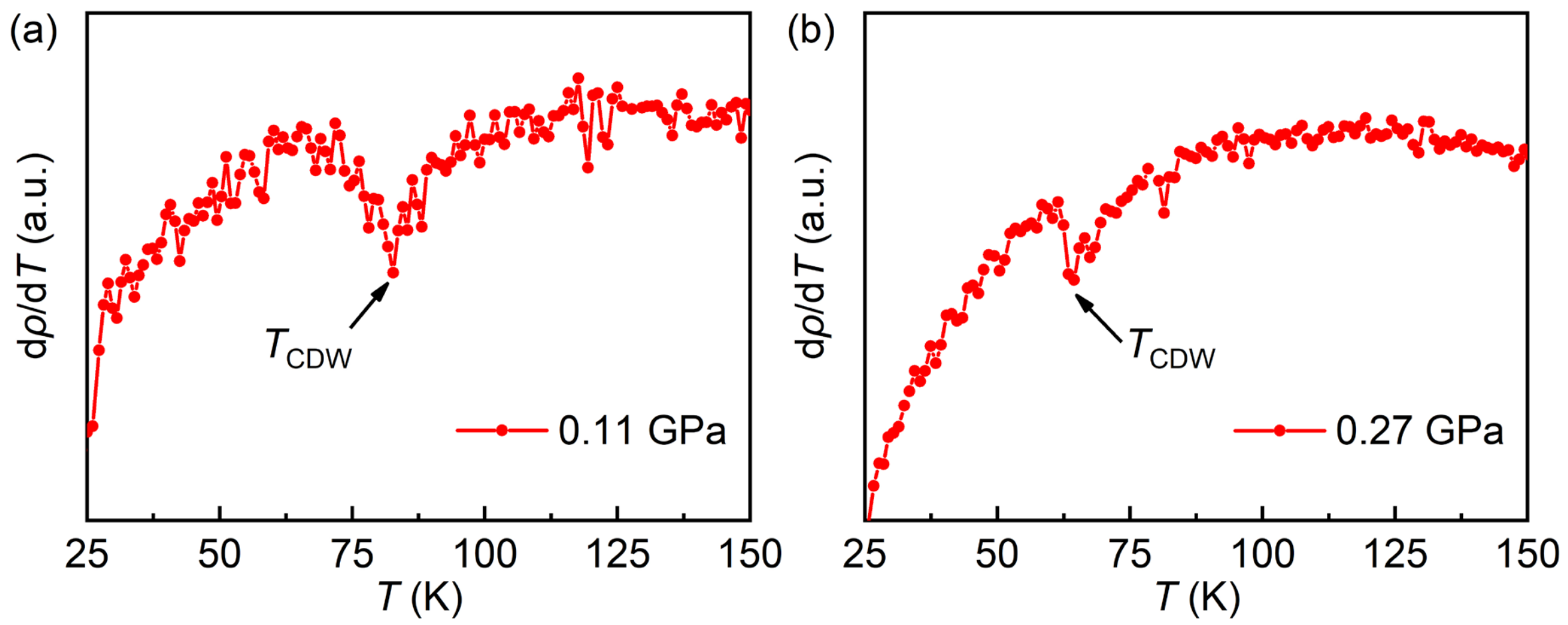

**Figure S4.** a-b) d$\rho$/d$T$ vs $T$ curves at 0.11 GPa and 0.27 GPa .

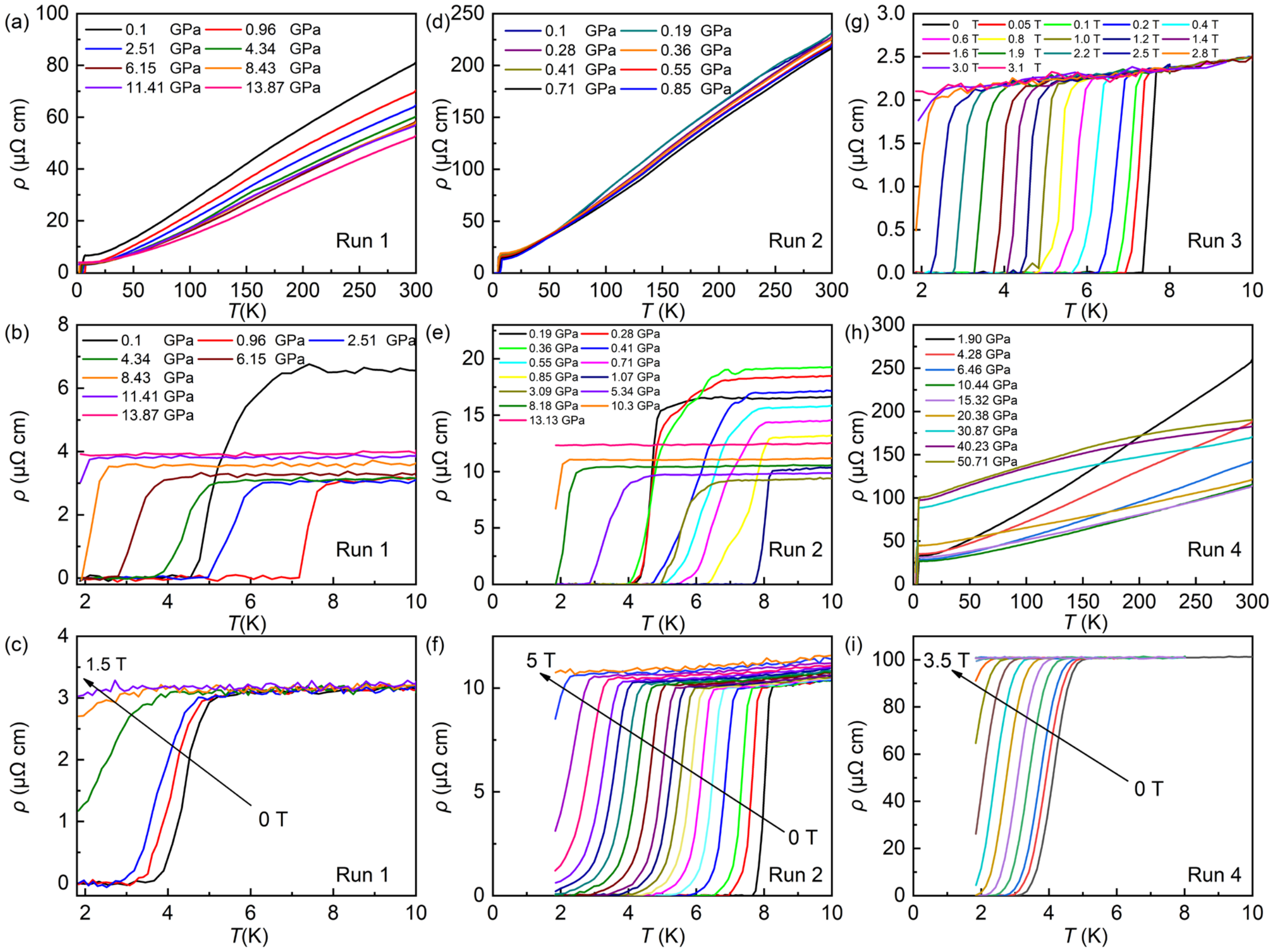

**Figure S5.** a) Resistivity $\rho(T)$ as a function of temperature under various pressures in Run 1 using NaCl as PTM. b) Enlarged part of $\rho(T)$ at low temperature region in Run 1. c) Temperature dependence of $\rho(T)$ at 4.34 GPa with the applied magnetic fields in Run 1. d) Temperature dependence of $\rho(T)$ at various pressures in Run 2 using NaCl as PTM. e) Enlarged part of $\rho(T)$ at low temperature range in Run 2. $T_{c,onset}$ reaches a maximum value ~ 8.2 K at 1.07 GPa. f) Temperature dependence of $\rho(T)$ under various fields at 1.07 GPa in run 2 using NaCl as PTM. g) Temperature dependence of $\rho(T)$ at 1.18 GPa at various magnetic fields in Run 3 using 7373 as PTM. h) Temperature dependence of $\rho(T)$ at various pressures in Run 4 using NaCl as PTM. i) Temperature dependence of $\rho(T)$ at 50.71 GPa at various magnetic fields in Run 4.

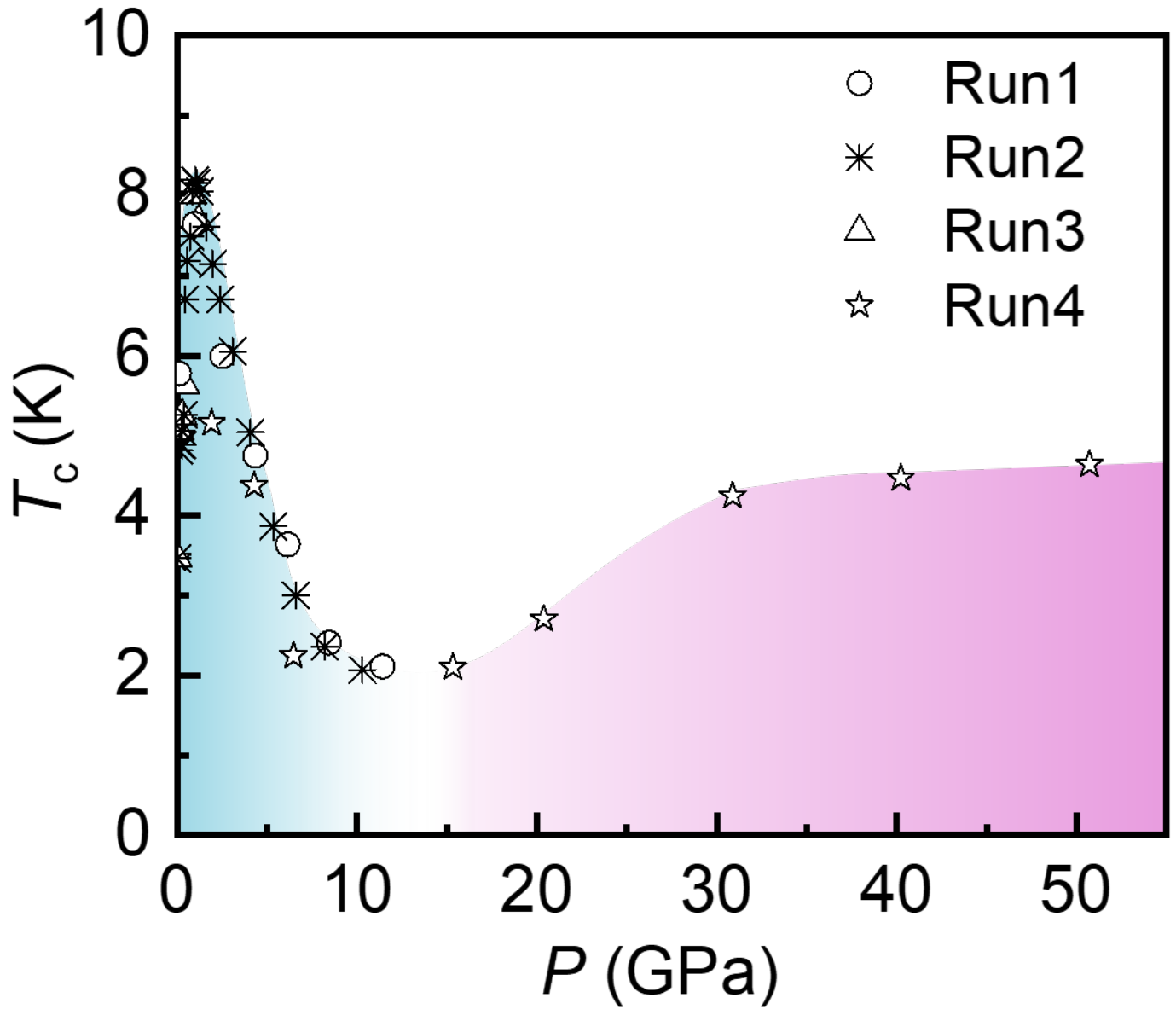

**Figure S6.** The temperature-pressure phase diagram of $CsV_3Sb_5$ up to 50.71 GPa in different runs.

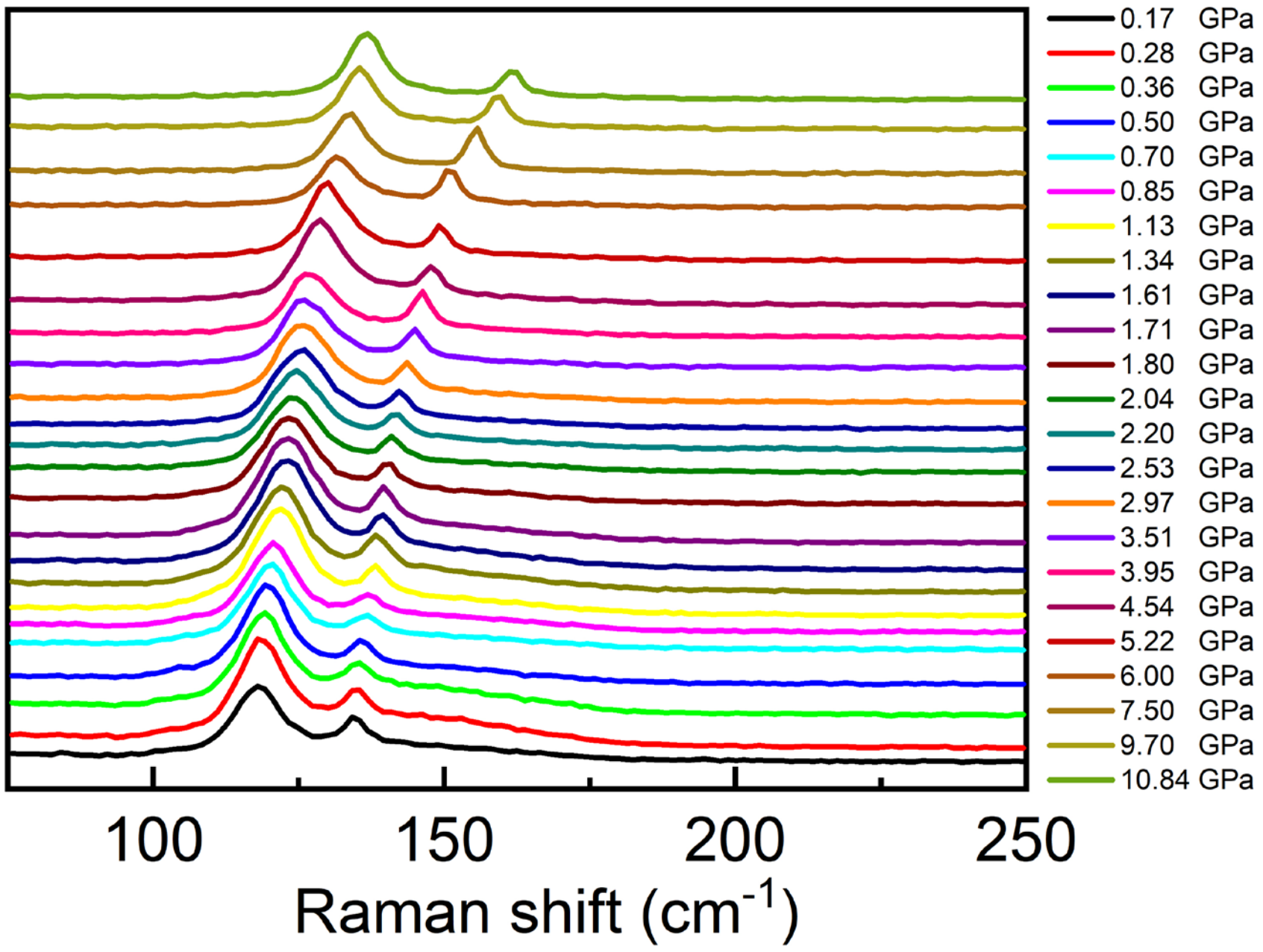

**Figure S7.** Raman spectra of $CsV_3Sb_5$ at 300 K under various pressures.